\documentclass[titlepage,11pt,twocolumn,letter]{article}

\input{psfig.sty}
\usepackage{isolatin}

\usepackage{graphicx}

\usepackage{ocg}
\usepackage{times}

\begin{document}

\title{Technical Report RT/46/2008\\\Huge{Optimizing Binary Code Produced by Valgrind}
\Large{(Project Report on Virtual Execution Environments Course -
AVExe)}}

\author{Filipe Cabecinhas~~~~~~~~~~~~~~~~Nuno Lopes~~~~~~~~~~~~~~~~Renato
Crisostomo
\\\\
Luis Veiga (\emph{Instructor})
\\\\
INESC-ID/IST, Distributed Systems Group, Rua Alves Redol N. 9,1000-029 Lisboa, Portugal\\
\{filipe.cabecinhas,nuno.lopes,renato.crisostomo,luis.veiga\}@ist.utl.pt}

\date{\Large{Aug 2008}
\begin{figure}[t]
\includegraphics[scale=0.3,width=7.5cm]{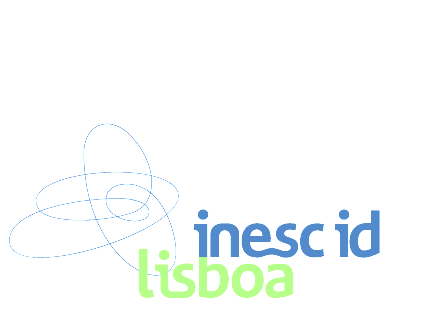}
\end{figure}
\newpage}
\maketitle

\begin{abstract}

Valgrind is a widely used framework for dynamic binary
instrumentation and it's mostly known by its memcheck tool.
Valgrind's code generation module is far from producing optimal
code. In addition it has many backends for different CPU
architectures, which difficults code optimization in an architecture
independent way. Our work focused on identifying sub-optimal code
produced by Valgrind and optimizing it.
\end{abstract}
\newpage\clearpage.\clearpage

%%%%%%%%%%%%%%%%%%%%%%%%%%%%%%%%%%%%%%%%%%%%%%%%%%%%%%%%%%%%%%%%%%%%%%%%%%%%%%%%%%%%%%%

\pagestyle{plain} \small
\section{INTRODUCTION}

Valgrind \cite{valgrind2007} is a widely used framework for
heavyweight dynamic binary instrumentation. It is mostly known by
its memcheck tool \cite{memcheck2005}, that is able to detect
several kinds of memory access errors.

Our work focused on identifying sub-optimal code produced by
Valgrind and optimizing it. Our optimizations were done on the
assembly level, rather than on the IR level. We decided not to do
them at the IR level because Valgrind already performs many
optimizations on this level and thus there would be fewer (not very
complex) things to do.

We begin this document by presenting a little introduction of
Valgrind's internals. We then describe our contributions, including
the problems we tried to fix and how they were fixed. Finally we
conclude with extensive results from benchmarks of each of the
proposed patches.

%%%%%%%%%%%%%%%%%%%%%%%%%%%%%%%%%%%%%%%%%%%%%%%%%%%%%%%%%%%%%%%%%%%%%%%%%%%%%%%%
\section{INTRODUCTION TO VALGRIND}

In this section we present a brief introduction of Valgrind's
internals. It is based on \cite{valgrind2007, NicholasThesis} and
the insight we acquired during this work.

Valgrind is a Process Virtual Machine that provides a framework for
building dynamic instrumentation tools. Memcheck \cite{memcheck2005}
is Valgrind's best known tool. It is able to detect several memory
related problems, like memory leaks, uninitialized memory positions
reads, invalid memory positions accesses, etc...

Valgrind is currently only used to do same-ISA virtualization,
although technically it could also provide virtualization for
different host and guest ISAs. Currently Valgrind officially
supports the following platforms: \{x86, amd64, ppc32, ppc64\}-linux
and \{ppc32, ppc64\}-AIX.

\subsection{The Core}
Valgrind's core is split in two: coregrind and VEX. VEX is
responsible for dynamic code translation and for calling tools'
hooks for IR instrumentation, while coregrind is responsible for the
rest (dispatching, scheduling, block cache management, symbol name
demangling, etc..). Our work has focused on VEX, with minor
incursions in coregrind.

\subsection{Code Translation}
Code translation is done by VEX and it is done in eight phases (as
of Valgrind 3.x). The phases are:

\begin{enumerate}
  \item Code disassembly: conversion of the machine dependent code to VEX's machine independent IR.
        The IR is based on single-static-form (SSA) and has some RISC-like features.
        Most instructions get disassembled to several IR opcodes.
  \item IR optimization: some standard compiler optimizations \cite{dragonBook} are applied
        to the IR, including
        dead code removal, constant folding, common sub-expression elimination (CSE), etc...
  \item Instrumentation: VEX calls the Valgrind tool's hooks to instrument the code.
  \item IR Optimization: similar to the previous optimization pass, albeit a little simpler.
  \item Tree Building: Transform the flat IR to tree IR, to simplify the next phase.
  \item Instruction Selection: conversion of the IR to machine code. This phase still uses virtual registers.
  \item Register Allocation: allocates real host registers to virtual registers, using
        a linear scan algorithm. This phase can create additional instructions for register
        spills and reloads (especially in register-constrained architectures like x86).
  \item Final code generation: generates the final machine code, by simply encoding the
        previously generated instructions and storing them to a memory block.
\end{enumerate}

At the end of each block, VEX saves all guest registers to memory,
so that the translation of each block is independent of the others.

\subsection{Block Management}
Blocks are produced by VEX, but are cached and executed by
coregrind. Each block of code is actually a superblock (single-entry
and multiple-exit).

Translated blocks are stored by coregrind in a big translation table
(that has a little less than 420,000 useful entries), so it rarely
gets full. The table is partitioned in 8 sectors. When it gets 80\%
full, all the blocks in a sector are flushed. The sectors are
managed in FIFO order. A comprehensive study of block cache
management can be found in \cite{cacheThesis}.

\subsection{Block Execution}
Blocks are executed through coregrind's dispatcher, a small
hand-written assembly loop (just 12 instructions on a x86-linux
platform). Each block returns to the dispatcher loop at the end of
its execution (and thus there's no block chaining).

The dispatcher lookups for blocks in a cache (with $2^{15}$
entries). The cache has an average hit-rate of about 98\%
\cite{valgrind2007}. When the cache lookup fails, the dispatcher
fallbacks to a slower routine written in C to lookup the translated
block in the translation table, or to translate the block if it
isn't in the table.

The dispatcher is also responsible for checking if the thread's
timeslice ended. When the timeslice ends, the dispatcher yields the
control back to the scheduler.

Valgrind has two dispatchers: normal (unprofiled), and profiled. The
profiled dispatcher is slower than the normal one, and is used to
gather statistic information (e.g. cache hit-rate).

\subsection{Assembly Conventions}
Throughout this document we will use x86 assembly to exemplify how a
certain optimization works. We will use Valgrind's assembly
conventions. Register references start with \texttt{\%}. Virtual
registers names are prefixed with \texttt{vr}, while others (real
host registers) don't have any prefix.

Take the following as an example:

\footnotesize\begin{verbatim}
1: movl %vr3, %eax
2: addl $1, %vr4
\end{verbatim}

The instruction in line 1 is a move from virtual register 3 to EAX
(a real host register). The instruction in line 2 sums the constant
\texttt{1} to the virtual register 4, and stores the result back to
the same register.

%%%%%%%%%%%%%%%%%%%%%%%%%%%%%%%%%%%%%%%%%%%%%%%%%%%%%%%%%%%%%%%%%%%%%%%%%%%%%%%%
\section{CONTRIBUTIONS}

In this section we present each of our contributions in detail. Each
one is provided as a patch against latest Valgrind SVN trunk at time
of writing (Valgrind: r8098; VEX: r1849). All patches were tested
with Valgrind's regression test suite and none of the patches
introduced a new failure. The patches were tested in four different
platforms: \{x86, amd64, ppc32, ppc64\}-linux.

\subsection{Peephole Optimizer}
VEX's code translation is currently split in eight phases, including
two optimization passes on the IR. We have created a new phase that
implements a simple peephole optimizer \cite{dragonBook}. The
peephole optimization phase is run after the instruction selection
phase and before the register allocation phase and thus operates on
non-register allocated (but ISA dependent) assembly code.

We have only implemented two simple optimizations on this phase,
although the creation of this phase leaves room for future peephole
optimizations (both architecture dependent and architecture
independent). Architecture independent optimizations are possible
because each backend implements a set of functions (e.g. isMove() or
getRegUsage()) that allows some high-level manipulation of the
ISA-dependent machine code.

The peephole optimizations that were implemented are: virtual to
virtual register coalescing and dead stores to virtual registers
elimination. The peephole optimizer starts by collecting virtual
register liveness information. Currently we only collect the last
time a register was used (either written or read). This information
is then used by both optimizations.

The virtual to virtual register coalescing pass eliminates redundant
MOVs. Take the following case as an example:
\footnotesize\begin{verbatim}
10: movl %vr3, %vr42
11: addl $1, %vr42
...
\end{verbatim}

If \texttt{\%vr3} isn't used after line 10, then we can simply
discard \texttt{\%vr42} and map it to \texttt{\%vr3}:
\footnotesize\begin{verbatim} ; previous line 10 was deleted
10: addl $1, %vr3
...
\end{verbatim}

Register coalescing is very important for SSA-based IR, like VEX's.
Unfortunately we only found that VEX's register allocator already
does this optimization after implementing it in the peephole pass.
It was an interesting exercise nevertheless.

The second peephole optimization was the elimination of dead stores
to virtual registers. It works by removing MOVs to virtual registers
that are never accessed again. This could be improved further if we
recorded finer-grained liveness information (at the expense of a
slower data collection phase), by eliminating MOVs whose value is
never read, e.g.: \footnotesize\begin{verbatim}
1: movl $2, %vr3
2: movl $1, %vr3
...
\end{verbatim}

Our implementation doesn't remove line 1 (clearly a dead store),
although the proposed one would. We believe that VEX's backends
don't produce such code, so it wouldn't be useful in practice.

Note that this second optimization could be ported to the register
allocator as well, eliminating the peephole pass altogether, and
thus possibly providing better performance. This wasn't done because
changing and tuning the register allocator is a very complex task
and we also wanted to prove the concept of a peephole optimizer for
valgrind.

The patch is named \texttt{vex\_peephole\_optimizations.txt} and is
independent of the host platform.

\subsection{Code Relocator}
VEX generates position-independent code, so that superblocks can be
easily moved around. Unfortunately this causes some sub-optimal x86
instructions to be emitted. We have identified two: absolute calls
and absolute jumps. The problem is that the x86 instruction set
doesn't have an instruction to do an absolute call/jump with an
immediate operand \cite{IntelInstructionSet}, although it has
instructions for relative call/jump with an immediate operand. So we
have implemented a code relocator that allows VEX to emit absolute
calls/jumps with a relative operand.

VEX's current implementation for absolute calls/jumps is to load the
address into a register and then do the jump/call based on the
register's value. E.g.:

\footnotesize\begin{verbatim}
movl $addr, %edx
jmp *%edx
\end{verbatim}

Our approach is to emit a single instruction (a relative jump/ call)
and thus save the extra two bytes. This is implemented in two parts:
in VEX and in coregrind. VEX was patched to emit relative
calls/jumps and to also provide a relocation table (an array with
the positions of the code that need to be patched when relocated).
Coregrind can then move the code to wherever needed and then call
the VEX relocator to patch the relative addresses to match their new
location.

%% Esta parte etá menos clara... Mas acho que escapa :-P
As a side effect, we save one register from spilling when calling
functions with four arguments (the maximum supported by VEX). For
functions with three or less arguments, VEX uses one of the caller
saved registers (per ABI convention). But as there are only three of
such registers, VEX must use an additional register when calling
functions with four arguments. For jumps, we also save one register
(\texttt{\%edx}).

To our best knowledge, code relocation isn't needed for PPC32 and
PPC64 architectures, as those architectures don't suffer from the
problem described before. We also believe that it is not possible to
port the patch to x86\_64 in a safe way, as there is no jump or call
instruction that takes a 64-bit immediate (either relative or
absolute) \cite{IntelInstructionSet}.

The patch is named \texttt{vex\_relocate\_abs\_calls.txt} and is
only implemented for x86 hosts.

\subsection{Instruction Pointer Store Optimization}
Valgrind often has to record the guest program's state, which
includes every register and flag in the processor. When recording
this state, the instruction pointer (IP) is frequently incremented
by a small amount. Our optimization stores only the least
significative bytes whenever possible with savings of up to 7 bytes
in amd64 and up to 12 bytes in PPC64 (biggest savings) in code size
per store. As the instruction pointer is often saved by Valgrind
(for example, to give meaningful error messages), with this
optimization, the code size becomes visibly smaller, which helps
reducing the program's cache misses and overall memory footprint.

We have implemented this optimization as follows: we track the
stores of the IP to memory, and we replace each store with a simpler
one that changes only the least significative bytes that were
changed since the last store. Often this mean storing only one or
two bytes.

Example: \footnotesize\begin{verbatim}
; PUT(60) = 0x80483D5:I32
movl $0x80483D5, 0x3C(%ebp)

; PUT(60) = 0x80483E8:I32
; note: x86 is little-endian
movb $0xE8, 0x3C(%ebp)
; instead of movl $0x80483E8, 0x3C(%ebp)
\end{verbatim}

This patch is architecture dependent and was split in two: one for
the Intel architectures (x86 and amd64), named
\texttt{vex-amd64-and-x86-IP-Store-optimization.txt}, and one for
the POWER architecture (PPC32 and PPC64), named
\texttt{vex-CIA-optimization.txt}.

\subsection{Dead Store to Real Register Elimination}
VEX's instruction selection pass sometimes produces virtual to real
register moves (e.g. when calling helper functions that receive the
arguments through registers). Our optimization eliminates these
instructions if the virtual register is mapped to the target (real)
register. This optimization was implemented in the register
allocator, by comparing the virtual register operand entry in the
register mapping table against the real register operand.

As an example, take the following x86 code:
\footnotesize\begin{verbatim}
movl %vr42, %eax
\end{verbatim}

If \texttt{\%vr42} is mapped to \texttt{\%eax}, the register
allocated code would become: \footnotesize\begin{verbatim}
movl %eax, %eax
\end{verbatim}

With our optimization, this instruction (a dead store) would be
discarded. This saves two bytes per each instruction removed in an
x86 host.

The patch is named \texttt{vex\_regalloc\_mov\_vr.txt} and is
independent of the host platform.

\subsection{Block Chaining}
Block chaining is a standard technique to improve VM's performance
\cite{VMBook}. Usually a superblock ends by jumping to the VM
dispatcher code, which introduces overhead in the execution and
messes up the CPU's branch prediction. Block chaining consists in
patching unconditional jump sites to do a direct jump to the target
superblock, bypassing the VM dispatcher.

Valgrind 2.x performed block chaining (briefly described in section
2.3.6 of \cite{NicholasThesis}), but Valgrind 3.x doesn't do it
(because it was a major rewrite and nobody implemented chaining
yet). It worked as follows: at the end of each superblock there's a
jump to the dispatcher, which gets patched by the dispatcher when
the target superblock address is known (i.e. when it is in cache).
Each superblock also has a prolog to check for thread timeslice end
and event checking. On a cache sector flush (managed in a FIFO way),
all blocks are scanned for patched jumps to flushed blocks, which
get unpatched (i.e. make them return to the dispatcher again). The
cost of scanning all blocks is high, but as it doesn't happen
frequently (because the block cache is big), this isn't a major
source of inefficiency.

We started to port Valgrind 2.4.1's block chaining code, but
unfortunately we didn't finish it. This code requires a code
relocator, so we had to implement it first (described previously)
and then we didn't have the time to finish this.

Although Valgrind's unprofiled (normal) dispatcher is faster than
many VMs' (it is just 12 instructions long on a x86-linux platform),
we believe block chaining should still give a good speedup (albeit
smaller than what other VMs have experienced
\cite{BlockChainingTR}).

The patch is named \texttt{valgrind\_block\_chaining.txt} and is
only implemented for x86 hosts. Although not complete, it's a good
starting point for future work.

\subsection{Misc}
In addition to the major contributions described in the previous
sections, we have also contributed minor patches to fix bugs found
during our work. We have provided patches for the following bugs
found:
\begin{itemize}
  \item some regression tests didn't compile on PPC64 due to a problem in a makefile, that was trying
        to link some PPC32 and PPC64 objects together. Patch name: \texttt{memcheck\_tests\_ppc64\_fix.txt}
        (patch already in Valgrind's official SVN tree).
  \item register liveness debug print on register allocator didn't compile.\\
        Patch name: \texttt{vex\_regalloc\_debug\_print\_fix.txt}
  \item in some cases the register allocator erroneously assumed that the opposite of a register write
        is a register read, which is not true, as VEX also has a modify access pattern (read plus write).
        The outcome was that some spills were skipped because it was assumed that the register value hadn't been modified.
        We were only able to observe this bug when using the Peephole optimizer (described previously).
        Marc-Oliver Straub also discovered this bug independently, so we assume the bug can be triggered without
        our Peephole optimizer.
        Patch name: \texttt{vex\_regalloc\_eqspill\_bugfix.txt}.\\
        Note: a similar patch was already committed to the official SVN repository.
  \item VEX couldn't emit an x86/amd64 instruction to store an immediate value in a memory location without
        using an additional register. This was needed to implement the instruction pointer store optimization
        in these architectures.
  \item at last, we have helped debugging the DRD tool on the PPC64-linux platform.
\end{itemize}

%%%%%%%%%%%%%%%%%%%%%%%%%%%%%%%%%%%%%%%%%%%%%%%%%%%%%%%%%%%%%%%%%%%%%%%%%%%%%%%%
\section{RESULTS}

In this section we present some experimental results of each of our
patches.

\subsection{Methodology}
To evaluate our contributions, we have run the standard Valgrind
performance tests (described in appendix \ref{appendixTests}). We
only present results for the memcheck tool, because of space
restrictions. However we provide raw results of the other tests
separately.

The tests were run in three different machines (where applicable):
\begin{itemize}
  \item Intel Pentium M 2.0 GHz (x86), 2 MB L2, 1 GB RAM
  \item AMD Athlon 64 3000+ 2.0 GHz (amd64), 512 KB L2, 1 GB RAM
  \item PlayStation 3, Cell 3.2 GHz (PPC64), 256 MB RAM
\end{itemize}

\subsection{Speedup}
In this section we present the speedup (in \%) achieved by each
optimization in each platform. The results presented are the mean of
three runs.

\begin{figure}[thpb]
  \centering
  \includegraphics[scale=0.2]{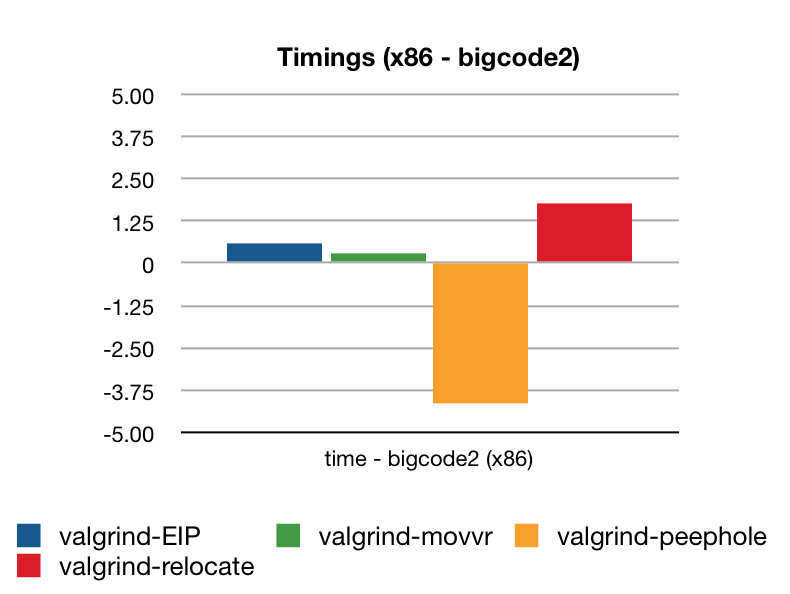}
  \caption{Speedup on the Bigcode 1 test on x86}
\end{figure}

\begin{figure}[thpb]
  \centering
  \includegraphics[scale=0.25]{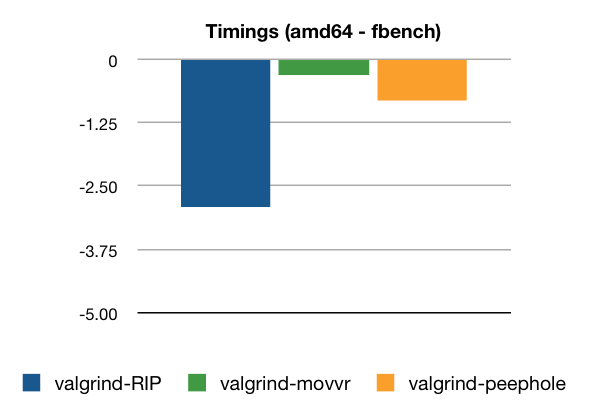}
  \caption{Speedup on the fbench test on amd64}
\end{figure}

\begin{figure}[thpb]
  \centering
  \includegraphics[scale=0.2]{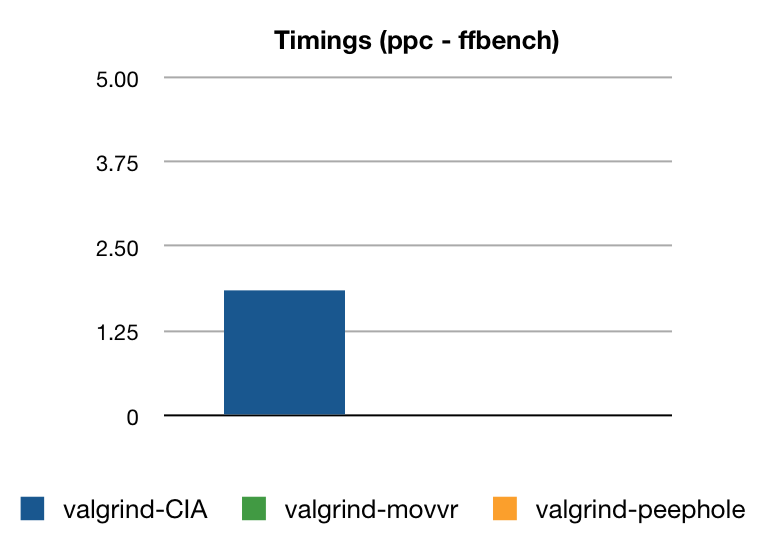}
  \caption{Speedup on the fbench test on PPC64}
\end{figure}

%\newpage
\subsection{Code Size Improvement}
In this section we present the changes (in \%) in the generated
machine code size. Note that higher values are better (i.e. smaller
code sizes).

\begin{figure}[thpb]
  \centering
  \includegraphics[scale=0.25]{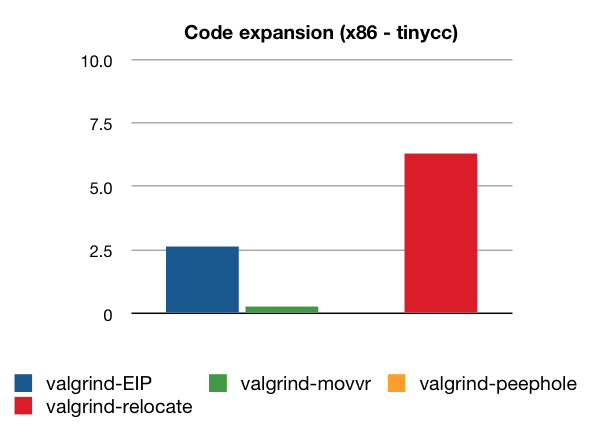}
  \caption{Code size reduction on the tinycc test on x86}
\end{figure}

\begin{figure}[thpb]
  \centering
  \includegraphics[scale=0.3]{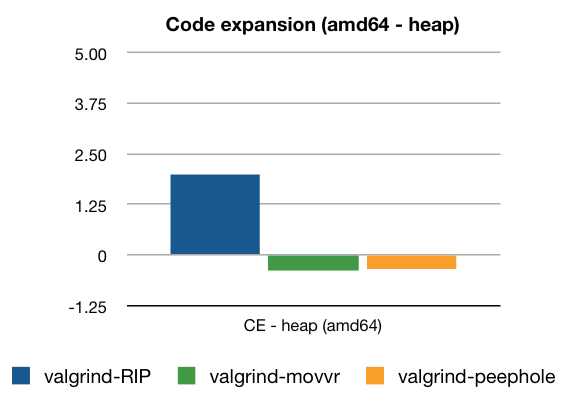}
  \caption{Code size reduction on the heap test on amd64}
\end{figure}

\begin{figure}[thpb]
  \centering
  \includegraphics[scale=0.3]{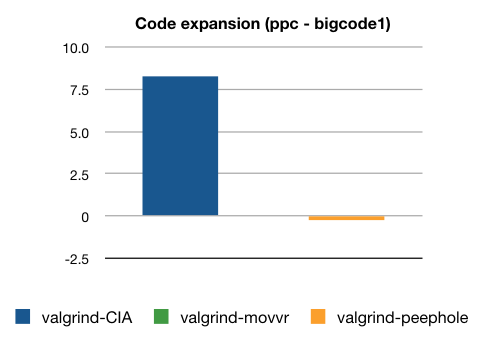}
  \caption{Code size reduction on the bigcode 1 test on PPC64}
\end{figure}

\subsection{Discussion}
All patches reduce the generated machine code size, which is great
for machines with less RAM and/or a small CPU cache. The EIP/RIP/CIA
optimization and the code relocator provide the most noticeable
results (i.e. they reduce the code size substantially). The
EIP/RIP/CIA optimization shows better results with code with many
memory accesses, as that's when VEX produces more stores of the
intruction pointer (as it has to bring the stored state up to date
before memory accesses). The code relocator consistently reduces the
code size of all tests.

One important thing to note is that code size reduction is
cumulative between tests. This means that applying more than one
patch will sum the code size reductions, as the patches optimize
different things.

Some patches also give noticeable speedup. Again this includes the
EIP/RIP/CIA optimization and the code relocator. The peephole
optimizer as-is doesn't provide a positive speedup, as it doesn't
feature many optimizations. The movvr optimization is neutral in
terms of speedup, but it should be considered for it's code size
reduction benefit.

%%%%%%%%%%%%%%%%%%%%%%%%%%%%%%%%%%%%%%%%%%%%%%%%%%%%%%%%%%%%%%%%%%%%%%%%%%%%%%%%
\section{CONCLUSIONS}

We have presented and implemented some optimizations in Valgrind
that reduce the size of the generated machine code, and give a
little speedup as well. These optimizations fix some problems we
have identified that led to Valgrind generating sub-optimal code.

Identifying the potential code for optimizations was actually quite
time consuming. Not only we knew nothing about Valgrind's internals,
reading machine generated low-level code and identifying
optimization opportunities is a very tricky job.

Other important thing to remember is that making optimizations for
JIT compilers is very difficult. This is because the optimization
cost must be amortized by the program running time, and thus many
optimizations that look great on the paper aren't useful in
practice.

We hope our patches can be integrated in a future Valgrind release.

%%%%%%%%%%%%%%%%%%%%%%%%%%%%%%%%%%%%%%%%%%%%%%%%%%%%%%%%%%%%%%%%%%%%%%%%%%%%%%%%
\section{FUTURE WORK}

Although Valgrind is a great tool and has a nice community always
trying to improve it, there's still some room for improvement.

We haven't finished implementing the block chaining optimization.
Finishing this task and porting the code to the other host
architectures is in our todo list. Selective unchaining
\cite{ShadrTR} is also worth investigating as it can reduce the size
of the prolog of the superblocks (and thus the overhead associated
with signal checking).

We believe that better register allocators may exist for JIT
environments like Valgrind (e.g. \cite{regAllocJIT}) other than the
linear scan algorithm used. Those algorithms usually also provide
better register coalescing than Valgrind's. Inter-block register
allocation (like Pin \cite{Pin} does) may also give a good speedup.

As usual, optimizing code is a never ending job, and a very
difficult one. Inspecting the code produced by Valgrind more
carefully (both the VEX IR and the assembly) may uncover other
potential optimizations that we have surely missed.

%%%%%%%%%%%%%%%%%%%%%%%%%%%%%%%%%%%%%%%%%%%%%%%%%%%%%%%%%%%%%%%%%%%%%%%%%%%%%%%%
\section{ACKNOWLEDGMENTS}
The authors thank Valgrind's team (especially Julian Seward and
Nicholas Nethercote) for answering our questions and for being so
supportive all the time. The authors also thank Prof. David Martins
de Matos for supplying a PPC64 machine for our tests and benchmarks.
Finally the authors would like to thank Prof. Luis Veiga for
providing ideas for this work and also for allowing us to work on
this project as a coursework.

%%%%%%%%%%%%%%%%%%%%%%%%%%%%%%%%%%%%%%%%%%%%%%%%%%%%%%%%%%%%%%%%%%%%%%%%%%%%%%%%

\bibliographystyle{plain}

%%%%%%%%%%%%%%%%%%%%%%%%%%%%%%%%%%%%%%%%%%%%%%%%%%%%%%%%%%%%%%%%%%%%%%%%%%%%%%%%
\onecolumn
\appendix
\section{Description of the Benchmark Tests}\vspace{-10pt}
\label{appendixTests} The following is copied from the perf/README
file found in the Valgrind source code.\vspace{-18pt}
\footnotesize\begin{verbatim} Artificial stress tests
-----------------------------------------------------------------------------
bigcode1, bigcode2: - Description: Executes a lot of (nonsensical)
code. - Strengths:   Demonstrates the cost of translation which is a
large part
               of runtime, particularly on larger programs.
- Weaknesses:  Highly artificial.

heap: - Description: Does a lot of heap allocation and deallocation,
and has a lot
               of heap blocks live while doing so.
- Strengths:   Stress test for an important sub-system; bug #105039
showed
               that inefficiencies in heap allocation can make a big
               difference to programs that allocate a lot.
- Weaknesses:  Highly artificial -- allocation pattern is not real,
and only
               a few different size allocations are used.

sarp: - Description: Does a lot of stack allocation and
deallocation. - Strengths:   Tests for a specific performance bug
that existed in 3.1.0 and
               all earlier versions.
- Weaknesses:  Highly artificial.

Real programs
-----------------------------------------------------------------------------
bz2: - Description: Burrows-Wheeler compression and decompression. -
Strengths:   A real, widely used program, very similar to the
256.bzip2
               SPEC2000 benchmark.  Not dominated by any code, the hottest
               55 blocks account for only 90% of execution.  Has lots of
               short blocks and stresses the memory system hard.
- Weaknesses:  None, really, it's a good benchmark.

fbench: - Description: Does some ray-tracing. - Strengths:
Moderately realistic program. - Weaknesses:  Dominated by sin and
cos, which are not widely used, and are
               hardware-supported on x86 but not on other platforms such as
               PPC.

ffbench: - Description: Does a Fast Fourier Transform (FFT). -
Strengths:   Tests common FP ops (mostly adding and multiplying
array
               elements), FFT is a very important operation.
- Weaknesses:  Dominated by the inner loop, which is quite long and
flatters
               Valgrind due to the small dispatcher overhead.

tinycc: - Description: A very small and fast C compiler.  A munged
version of
               Fabrice Bellard's TinyCC compiling itself multiple times.
- Strengths:   A real program, lots of code (top 100 blocks only
account for
               47% of execution), involves large irregular data structures
               (presumably, since it's a compiler).  Does lots of
               malloc/free calls and so changes that make a big improvement
               to perf/heap typically cause a small improvement.
- Weaknesses   None, really, it's a good benchmark.
\end{verbatim}

\end{document}